\documentclass[aps,twocolumn,prd,showpacs,nofootinbib]{revtex4}
\usepackage{amsmath}
\usepackage{graphicx}
\usepackage{dcolumn}
\usepackage{bm}
\usepackage{amssymb}
\usepackage{latexsym}

\def\be{\begin{equation}}
\def\ee{\end{equation}}
\def\ba{\begin{eqnarray}}
\def\ea{\end{eqnarray}}

\begin{document}

\title{On Dualities Of Primordial Perturbation Spectra }

\author{Yun-Song Piao$^{a,b}$}
\email{yspiao@itp.ac.cn}
\affiliation{${}^a$Institute of High Energy Physics, Chinese
Academy of Science, P.O. Box 918-4, Beijing 100039, P. R. China}
\affiliation{${}^b$Interdisciplinary Center of Theoretical
Studies, Chinese Academy of Sciences, P.O. Box 2735, Beijing
100080, China}


\begin{abstract}

In this paper, we discuss the dualities of the primordial
perturbation spectra from various expanding/contracting phases for
full space of parameter $w\equiv {p\over \rho}$ of state equation.

\end{abstract}

 \maketitle

The primordial perturbations on large scales may be an important
test for various scenarios of the early universe, thus it is very
significant to further probe their possible nature and origin. The
basic idea of inflation is simple and elegant \cite{GLS}, for a
recent review see \cite{LIN}. During inflation, the fluctuations
are stretched to the outside of horizon and form nearly
scale-invariant primordial perturbations leading to the formation
of cosmological structure \cite{MC}. Wands found \cite{W} (see
also \cite{FB}, for an earlier reference see \cite{S} ) that a
collapsing universe with $a\sim (-t)^{2\over 3}$ ( further for Pre
Big Bang-like scenario see \cite{GV, V} ), in which $a(t)$ is the
cosmological scale factor, can also generate such scale-invariant
spectrum, which is dual to the case of inflation with $a\sim
e^{ht}$ or $a\sim t^n, n\rightarrow +\infty$. This duality is a
result that the spectrum of the curvature perturbation $\zeta$ on
uniform comoving hypersurfaces is invariant under some
transformation. Recently, it has been noted \cite{GKS, KST} (see
also \cite{KOST, TT}) that instead of the curvature perturbation
$\zeta$, when the Bardeen potential $\Phi$ is considered, the
scale-invariant spectrum can be produced during a period of slow
contraction with $a\sim (-t)^n, n\rightarrow 0_+$ in cyclic
scenario \cite{KOS}, for some criticisms see \cite{FB, H, KKL,
TBF} . This agreement between physically dissimilar models with
different cosmological solutions is not coincidental. As has been
showed further by Boyle {\it et.al.} \cite{BST}, the relationship
between inflation and cyclic scenario can be regarded as a special
case of general duality of the perturbation spectrum of Bardeen
potential $\Phi$.

For the parameter $w\equiv {p\over \rho} \geq -1$ of state
equation, Boyle {\it et.al.}'s duality relates a stable expanding
solution to an stable/unstable contracting solution dependent on
the value of $w$, while Wands's duality relates a stable expanding
solution $(-1<w<-1/3)$ to an unstable contracting solution
$(-1/3<w<0)$, but that of $w>0$ has not relevant dual branch, thus
in some sense the dualities of perturbation spectrum may be
uncomplete when restricted to $w>-1$. When relaxing this limit,
the nearly scale-invariant perturbation spectrum can be also
similarly obtained during slow expansion \cite{PZ} and phantom
inflation \cite{PZ1}, but with different cosmological background
evolutions $a\sim (-t)^{n}$, in which $n\rightarrow 0_-$ and
$n\rightarrow -\infty$ respectively. In this brief report, we
discuss the dualities of the primordial perturbation spectra for
full space of constant $w$, which can be implemented in the frame
of single normal/phantom scalar field \cite{PZ2}.

In general the evolution of cosmological scale factor before the
``bounce" \footnote{ Here the ``bounce" means the exit from
pre-bounce expanding/contracting phases to late-time observational
cosmology, which corresponds to the usual reheating \cite{KLS,
FKL} for inflation scenario.}
in Einstein frame can be written as \be a(t)\sim t^n \label{at}\ee
in which $t\rightarrow +\infty$, or \be a(t)\sim (-t)^n
\label{ant}\ee in which $t\rightarrow 0_-$, $n$ is a positive or
negative constant. For $n>0$, Eq. (\ref{at}) corresponds to the
expanding phase, in which the perturbation spectrum was studied
firstly in Ref. \cite{AW}, and Eq. (\ref{ant}) corresponds to the
contracting phase. For $n<0$ the case is in reverse. The Fridmann
equations are \be h^2\equiv ({{\dot a}\over a})^2={\rho \over
3}\label{h} \ee \be {\dot h}= - {\rho +p\over 2}\label{doth} \ee
where $h$ is Hubble parameter and $8\pi G =1$ is set. For the case
that the speed of sound $c_s^2$ is constant, causally primordial
perturbations can be generated in such a phase, in which the
fluctuations exits the horizon and then re-enters the horizon
after the ``bounce" to an expanding phase corresponding to our
observational cosmology. This requires that $ah$ increases with
time, thus $n>1$ for (\ref{at}) and $n<1$ for (\ref{ant}) must be
satisfied. From (\ref{at}), (\ref{ant}) and (\ref{h}), we have \be
h={n\over t} \ee for various phases. We plot the sketch of
relations between $h$ and $t, n$ in Fig.1. These various
expanding/contracting phases can be implemented by the evolution
of single normal/phantom scalar field with exponential potential,
in this case $c_s^2 =1$ \cite{GM}, which is summarized in Table I.

\begin{figure}[]
\begin{center}
\includegraphics[width=8cm]{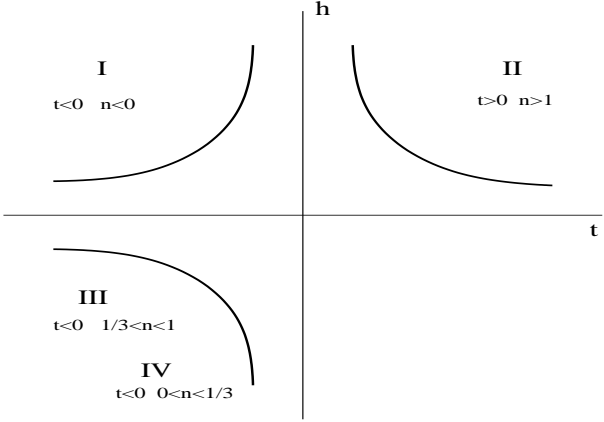}
\caption{The sketch of relations between Hubble parameter $h$ and
$t, n$: Various expanding/contracting phases are denoted by Region
I-IV in Table I. } \label{fig1}
\end{center}
\end{figure}

\begingroup
\begin{table*}
\begin{tabular}{||p{3.5cm}|p{2.5cm}|p{2.8cm}|p{2.5cm}|p{2.5cm}||}\hline \hline
Region & I & II & III & IV \\
\hline Evolution of scale factor & $a(t)\sim (-t)^n$ & $a(t)\sim t^n$ & $a(t)\sim (-t)^n$ & $a(t)\sim (-t)^n$ \\
\hline  & $n<0$ & $n>1$ & ${1\over 3}<n<1$ & $0<n<{1\over 3}$ \\
\hline Hubble parameter & $h>0, {\dot h}>0$ & $h>0, {\dot
h}<0$ & $h<0, {\dot h} <0$ & $h<0, {\dot h}<0$ \\
\hline & expansion & expansion & contraction & contraction \\
\hline Stability & stable & stable & unstable & stable \\
\hline Parameter of state equation & $w <-1$ & $-1<w <-{1\over 3}$ & $-{1\over 3} < w <1$ & $w >1$ \\
\hline $\epsilon$ & $\epsilon <0$ & $0<\epsilon <1$ & $1<\epsilon
<3$ & $\epsilon >3$ \\
\hline Kinetic energy term & reverse (phantom)
& standard & standard & standard \\
\hline
Potential energy term & standard & standard & standard & reverse (negative potential) \\
\hline $\Phi$ invariant dual regions & I & III or IV & II & II \\
\hline $\zeta$ invariant dual regions & III & III & I or II or IV
& III
\\ \hline \hline
\end{tabular}
\caption{The natures of various expanding and contracting phases
generating the primordial perturbation spectra: These phases can
be implemented by simple (normal/phantom) scalar field. Some
further details can be seen in our paper \cite{PZ2}. }
\end{table*}
\endgroup

Combing (\ref{at}), (\ref{ant}), (\ref{h}) and (\ref{doth}), the
power-law index of scale factor is given by \be n={2\over
3(1+w)}\label{w}\ee Instead of $w$, it may be more convenient to
parameterize the state equation with \be \epsilon \equiv {3\over
2}(1+w)\equiv {1\over n}\label{epsi}\ee For $w\simeq -1$,
$\epsilon\simeq 0$ is the usual slow-roll parameter of inflation
models.

In the following we briefly review the primordial perturbation
spectra from these expanding/contracting phases before the
``bounce". Let us pay attention to the scalar metric fluctuations.
In longitudinal gauge and in absence of anisotropic stresses, the
scalar metric perturbation can be written as \be ds^2 = a^2(\eta)
(-(1+2\Phi)d\eta^2 +(1-2\Phi) \delta_{ij}dx^i dx^j )\ee where
$\eta$ is conformal time $d\eta \equiv{dt\over a}$, thus \be
-\eta\sim (\pm t)^{-n+1} \label{eta} \ee \be a(\eta) \sim
(-\eta)^{{n\over 1-n}}\equiv (-\eta)^{1\over \epsilon -1}
\label{aeta}\ee and $\Phi$ is the Bardeen potential. The curvature
perturbation on uniform comoving hypersurfaces is given by \be
\zeta = {1\over \epsilon}({a\over a^\prime} \Phi^\prime+\Phi)+\Phi
\ee


After defining new variables \cite{L, M} (for a thorough
introduction to gauge-invariant perturbations see \cite{KS, MFB}),
\be u= {a\over \sqrt{2|{\cal H}^2-{\cal H}^\prime |}}\Phi ={a\over
\varphi^\prime}\Phi \ee \be v= {a\sqrt{2|{\cal H}^2 -{\cal
H}^\prime |}\over {\cal H}} \zeta ={a\varphi^\prime\over {\cal H}}
\zeta \ee where ${\cal H}\equiv {a^\prime \over a}$, the
perturbation equations of $\Phi$ and $\zeta$ for simple
(normal/phantom) scalar field can be written as \be
u_k^{\prime\prime}+\left(k^2 - {\mu^2-{1\over 4}\over
\eta^2}\right)u_k =0 \label{uk}\ee \be
v_k^{\prime\prime}+\left(k^2 -{\nu^2 -{1\over 4}\over
\eta^2}\right)v_k =0 \label{vk}\ee For all interesting modes $k$,
we can solve Eqs. (\ref{uk}) and (\ref{vk}) analytically and
obtain \be u_k = \sqrt{-k\eta}\left( B_1(k)J_{\mu }
(-k\eta)+B_2(k)J_{-\mu }(-k\eta)\right)\label{uks}\ee \be v_k =
\sqrt{-k\eta}\left( C_1(k)J_{\nu}
(-k\eta)+C_2(k)J_{-\nu}(-k\eta)\right)\label{uks}\ee where \be
\mu={1\over 2}|{\epsilon +1\over \epsilon -1 }|~,~~~~ \nu =
{1\over 2} | {\epsilon -3\over \epsilon -1}|~,\label{mu}\ee $J$ is
the first kind of the Bessel function with order $\mu$ or $\nu$,
the function $B_i(k)$ and $C_i(k)$ can be determined by specifying
the initial conditions.

In the regime $k\eta \rightarrow \infty $, in which the mode $u_k$
and $v_k$ are very deep in the horizon, Eqs. (\ref{uk}) and
(\ref{vk}) are reduced to the equations for a simple harmonic
oscillator, in which $ u_k \sim {e^{-ik\eta} \over (2k)^{3/2}}$
and $v_k \sim {e^{-ik\eta} \over (2k)^{1/2}}$ are stable. In the
regime $k\eta \rightarrow 0$, in which the mode $u_k$ and $v_k$
are far out the horizon, the modes are unstable and grows. In
long-wave limit, $\Phi_k$ and $\zeta_k$ can be given and expanded
to the leading term of $k$, \be k^{3\over 2} \Phi \sim k^{{1\over
2}-\mu}\label{Phi}\ee \be k^{3\over 2} \zeta \sim k^{{3\over
2}-\nu} \label{zeta} \ee Thus the spectrum indexes are \be
n_{\Phi}-1 =1-2\mu= 1- |{\epsilon +1\over \epsilon -1
}|\label{nphi}\ee \be n_{\zeta} -1=3- 2\nu = 3- | {\epsilon
-3\over \epsilon -1}|\label{nzeta}\ee

From Eqs. (\ref{nphi}) and (\ref{nzeta}), we can see that the the
spectrum index of $\zeta$ is the same as of $\Phi$ only for
$-1<\epsilon < 1$ \footnote{ It is known that the comoving
curvature perturbation $\zeta$ and the Bardeen potential $\Phi$
have different spectral dependence during the collapse and it is
impossible for both them to obey non-singular evolution equations
through the bounce \cite{CDC}, see also Ref. \cite{FB, PP, MP,
AWA} and Ref. \cite{PZ2} for a relevant discussions.  Which of the
spectra of $\Phi$ and $\zeta$ can be inherited in late-time
observational cosmology dependent on the matching conditions
through the ``bounce", {\it i.e.} how $\Phi$ and $\zeta$ pass
through the ``bounce", which is determined by the details of
``bouncing" physics. }, and when $\epsilon\simeq 0$, nearly
scale-invariant can be obtained, which corresponds to the case in
inflationary cosmology, but for other value of $\epsilon$, the
spectrum index of $\zeta$ is different from that of $\Phi$, the
$\Phi$ spectrum is nearly scale-invariant for $\epsilon
\rightarrow \pm \infty$, while the $\zeta$ spectrum is nearly
scale-invariant for $\epsilon \simeq {3\over 2}$.

Further, notice from Eq. (\ref{mu}) that $\mu$ is invariant under
$ {\epsilon +1\over \epsilon -1 }\rightarrow -{\epsilon +1\over
\epsilon -1 } $, which is equal to $\epsilon \rightarrow {1\over
\epsilon}$, in which the fixed points are $\epsilon=\pm 1$. Thus
the spectrum index of $\Phi$ is invariant under this
transformation, which can be also seen from Eqs. (\ref{Phi}) and
(\ref{nphi}). Similarly, the spectrum index of $\zeta$ is
invariant under the transformation of ${\epsilon -3\over \epsilon
-1}\rightarrow -{\epsilon -3\over \epsilon -1}$, {\it i.e.}
$\epsilon \rightarrow {2\epsilon-3\over \epsilon -2}$, in which
the fixed points are $\epsilon =1,3$.

\begin{figure}[t]
\begin{center}
\includegraphics[width=8cm]{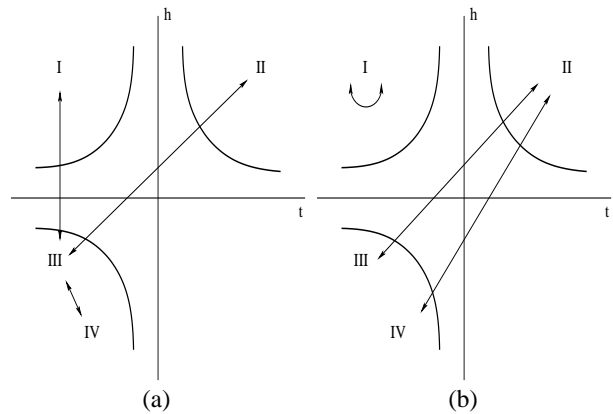}
\caption{The sketch of general dualities of $\zeta$(a) or
$\Phi$(b) spectrum: Various expanding/contracting phases are
denoted by Region I-IV in Table I. For $\zeta$ spectrum, stable
Regions I, II, IV are dual to unstable Region III, and for $\Phi$
spectrum, stable Region I is self-dual and stable Region II is
dual to unstable Region III and stable Region IV. } \label{fig2}
\end{center}
\end{figure}

\begingroup
\squeezetable
\begin{table}

    \label{spectrum}
    \begin{tabular}{|c|c|c|c|c|}
      \hline
      \multicolumn{5}{|c|}{The Dualities of Nearly Scale-Invariant Spectra }
       \\ \hline \hline  & \multicolumn{2}{|c|}{$\Phi$ spectrum } &
       \multicolumn{2}{|c|}{$\zeta$ spectrum } \\
      \hline \hline
      Phase & (Phantom) Inflation
      & Inflation  &
      \multicolumn{2}{|c|}{ Inflation} \\
      & $(a\sim (-t)^{n}, n\rightarrow -\infty)$ & $ (a\sim t^n, n\rightarrow
      +\infty)$ & \multicolumn{2}{|c|}{} \\
      \hline  \hline
      Dual & Slow Expansion & Slow
      Contraction & \multicolumn{2}{|c|}{Contraction }\\
     Phase &  $(a\sim (-t)^{n}, n\rightarrow 0_-)$ &  $(a\sim (-t)^{n}, n\rightarrow 0_+)$ &
      \multicolumn{2}{|c|}{ $(a\sim (-t)^{2\over 3} )$}\\
      \hline \hline
      Ref. & & \cite{GKS, KST, BST} &
      \multicolumn{2}{|c|}{\cite{W, FB, S}} \\ \hline \hline
    \end{tabular}
    \caption{The dualities of nearly scale-invariant spectra: For
$\Phi$ spectrum, the slowly expanding phase with $\epsilon
\rightarrow -\infty$ (The relation between $\epsilon$ and $n$ can
be seen from Eq. (\ref{epsi})) is dual to the phantom inflationary
phase with $\epsilon = 0_-$, in which the nearly scale-invariant
blue spectrum is given, and the slowly contracting phase with
$\epsilon\rightarrow \infty$ is dual to the usual inflationary
phase with $\epsilon = 0_+$, in which the nearly scale-invariant
red spectrum is given. For $\zeta$ spectrum, the inflationary
phase with $\epsilon\simeq 0$ is dual to the unstable contracting
phase with $\epsilon\simeq {3\over 2}$. }
  \end{table}
  \endgroup

Therefore, for $\epsilon$ being in the range $0\leq\epsilon
<\infty$, as has been shown in Ref. \cite{BST}, Wands's duality
pairs the expanding phase $(0\leq\epsilon <1)$ to the contracting
phase $(1<\epsilon \leq {3\over 2})$, but the contracting phase
with $\epsilon \geq 3$ is paired the contracting phase
$(2<\epsilon\leq 3)$ and have no relevant expanding dual branch,
while Boyle {\it et.al.}'s duality pairs the expanding phase
$(0\leq \epsilon<1)$ to the contracting phase $(1<\epsilon
<\infty)$. Further, when relaxed $\epsilon$ to $\epsilon <0$, the
dualities from various phases may be described more completely and
can be regarded as the complementarity of Ref. \cite{BST}. In this
case, Wands's duality pairs the expanding phase $(-\infty<\epsilon
<0)$ to the contracting phase $({3\over 2}<\epsilon <2)$, and
Boyle {\it et.al.}'s duality pairs the slowly expanding phase
$(-\infty<\epsilon\leq -1)$ to the expanding phase
$(-1\leq\epsilon <0)$.

The scale solution is a stable attractor for the expanding phase
if and only if $w<1$ $(\epsilon <3) $, and for the contracting
phase if and only if $w>1$ $(\epsilon >3)$ \cite{GKS, EWST}. Thus
Wands's duality relates the stable expanding/contracting solutions
to the unstable contracting solutions, while Boyle {\it et.al.}'s
duality relates two stable branches only when $\epsilon<1/3$ or
$\epsilon>3$, in terms of the spectrum index, which requires
$n_{\Phi,\zeta}
>0$. for $1/3<\epsilon<3$ the stable expanding solution is related
to the unstable contracting branch, in this case $n_{\Phi,\zeta}
<0$.

Due to these dualities, the primordial scalar perturbation spectra
may be not used to determine in which phase the perturbations are
generated. However, this degenerations may be remove by the tensor
perturbation spectrum, whose spectrum index is \cite{BST} \be n_t=
3-|{\epsilon-3\over \epsilon -1}|\ee The phase with $h$ rapidly
increasing will produce a more blue tensor spectrum than its dual
phase with $h$ slowly increasing or decreasing. The figure of the
tensor perturbation spectrum with $w$ can be seen in Ref.
\cite{PZ2}. For $\zeta$ spectrum, its scalar spectrum index is the
same as that of tensor, but fortunately, in this case, for
arbitrary stable phase, its stable dual branch does not exist. As
a result if the spectrum is generated by a stable solution then
the detection of the primordial gravitational wave background can
give a direct record of the evolution of the scale factor, thus a
direct selection for different expanding/contracting phases.

In summary, we discuss the dualities of the primordial
perturbation spectra for full space of constant $w$ (thus
$\epsilon$). The results are consistent with Boyle {\it et.al.}'s
for the case of $\epsilon
>0$ \cite{BST}. The exist of these dualities means that various cosmological
background solutions are dual each other in the sense of scalar
perturbation spectrum, see Fig.2 for a sketch of dualities of
various phases. Further, the nearly scale-invariant spectrum can
be from various cosmological scenarios is not also coincidental,
the relationship between them can be regarded as some special
cases of general dualities of $\Phi$ or $\zeta$ spectrum, see
Table II for a summary on the dualities of nearly scale-invariant
spectra, which may provide a possibility that the perturbations of
background solutions with $|\epsilon | >1$ can be studied by using
slow-roll $(|\epsilon | <1)$ method \cite{GKS, KST2}. In this
brief report, we only focus on the case of the single
(normal/phantom) scalar field with exponential potential in which
$w$ is constant and $c_s^2 =1$. The studies of relaxing these
limits may be more interesting.

\textbf{Acknowledgments} The author would like to thank Paul J.
Steinhardt for helpful discussions and comments on the manuscript.
This work is supported by K.C.Wang Postdoc Foundation.

\end{document}